\documentclass{pasj00}
\begin{document}
\SetRunningHead{T. Kato et al.}{SU UMa-Type Dwarf Nova WX Ceti}

\Received{2001/6/11}%{yyyy/mm/dd}
\Accepted{2001/7/11}%{yyyy/mm/dd}

\title{1998 Superoutburst of the Large-Amplitude\\SU~UMa-Type Dwarf Nova
WX Ceti}

\author{Taichi \textsc{Kato}, and Katsura \textsc{Matsumoto}}
\affil{Department of Astronomy, Kyoto University, Sakyo-ku, Kyoto 606-8502}
\email{tkato@kusastro.kyoto-u.ac.jp, katsura@kusastro.kyoto-u.ac.jp}

\author{Daisaku \textsc{Nogami}}
\affil{Hida Observatory, Kyoto University, Kamitakara, Gifu 506-1314}
\email{nogami@kwasan.kyoto-u.ac.jp}

\author{Koichi \textsc{Morikawa}}
\affil{468-3 Satoyamada, Yakage-cho, Oda-gun, Okayama 714-1213}
\email{koichi@morikawa.org}

\email{\rm{and}}

\author{Seiichiro \textsc{Kiyota}}
\affil{Variable Star Observers League in Japan (VSOLJ),
       1-401-810 Azuma, Tsukuba, Ibaraki, 305-0031}
\email{skiyota@nias.affrc.go.jp}

%%% end:list of authors


\maketitle

\begin{abstract}
   We observed the 1998 November superoutburst of WX Cet, a dwarf nova
originally proposed as a WZ Sge-like system.  The observation established
that WX Cet is an SU UMa-type dwarf nova with a mean superhump
period of 0.05949(1) d, which is 2.1\% longer than the reported orbital
period.  The lack of early superhumps at the earliest stage of the
superoutburst, the rapid development of usual superhumps, and the possible
rapid decay of late superhumps seem to support that WX Cet is a fairly normal
large-amplitude SU UMa-type dwarf nova, rather than a WZ Sge-type dwarf
nova with a number of peculiarities.  However, a period increase of
superhumps at a rate $\dot{P}/P$ = +8.5$\pm$1.0 $\times$ 10$^{-5}$ was
observed, which is one of the largest $\dot{P}/P$ ever observed in
SU UMa-type dwarf novae.  A linear decline of light, with a rate of 0.10
mag d$^{-1}$, was observed in the post-superoutburst stage.  This may be an
exemplification of the decay of the viscosity in the accretion disk after
the termination of a superoutburst, mechanism of which is proposed to
explain a variety of post-superoutburst phenomena in some SU UMa-type
dwarf novae.
\end{abstract}

\section{Introduction}
   Dwarf novae are a class of cataclysmic variables (CVs), which are
close binary systems consisting of a white dwarf and a red dwarf secondary
transferring matter via the Roche lobe overflow.  The resultant accretion
disk around the white dwarf is susceptible to various kinds of
instabilities, and is a source of a rich variety of activities in CVs.
The two most relevant instabilities are thermal and tidal instabilities,
which are responsible for dwarf nova-type outbursts and superhumps,
respectively [see \citet{osa96} for a review].  Systems having both
low mass-transfer rates ($\dot{M}$) and low mass ratios ($q=M_2/M_1$)
are susceptible to both thermal and tidal instabilities, and are called
SU UMa-type dwarf novae [for a recent review of SU UMa-type stars and their
observational properties, see \citet{war95}].  Among SU UMa-type dwarf
novae, there exists a small subgroup of WZ Sge-type dwarf novae
[originally proposed by \citet{bai79}; see also \citet{dow81} and
\citet{odo91}].  \citet{bai79} lists two dwarf novae, WX Cet and UZ Boo,
which show very infrequent and large-amplitude outbursts, similar to
those observed in WZ Sge.  WZ Sge, itself, has a number of peculiar properties
(e.g. the longest observed recurrence time of 33 years and the lack of
normal outbursts) among dwarf novae.  The WZ Sge-type phenomenon has been
a long-standing problem, and several hypotheses have been proposed.
Two historically representative ones are \citet{osa95} and \citet{las95};
the former assumed extremely low quiescent viscosity, and the latter assumed
the evaporation of the inner disk.  Modern views on these theoretical
models can be found in \citet{mey98} and \citet{min98}.  Since WX Cet was
selected as one of two relatives to WZ Sge by \citet{bai79}, this star
has been receiving much attention.

   \citet{odo91} observed the 1989 June superoutburst of WX Cet, and
detected superhumps with a period of $\sim$80 m.  \citet{odo91} argued,
from a similarity of photometric and spectroscopic features of WX Cet
to other SU UMa-type dwarf novae, that there is no clear reason to
retain the distinction between WZ Sge-type stars and other SU UMa-type
dwarf novae.  However, even if the argument by \citet{odo91} against the
clear observational separation of WZ Sge-type dwarf novae and SU UMa-type
dwarf novae is partly supported by recent observations of other stars
(e.g. \cite{nog96}; \cite{bab00}), the 1989 June superoutburst of WX Cet
was observed only under unfavorable seasonal condition, which made any
detailed observation and analysis difficult.  \citet{men94} performed
a radial-velocity study in quiescence, and obtained a most probable orbital
period ($P_{\rm orb}$) of 79.16(4) min.  \citet{kat95} performed
CCD photometry of the 1991 superoutburst, which again occurred in an
unfavorable seasonal condition, and found that the rate of decline
remarkably slowed down during the latter course of the superoutburst.
\citet{kat95} suggested some similarities to superoutbursts of WZ Sge,
itself.  More recently, \citet{rog01} observed WX Cet in quiescence,
and obtained a photometric period of 0.05827(2) d, which is in good
agreement with the spectroscopic orbital period by \citet{men94}.
\citet{rog01} further noted that the orbital light curve showed
alternations between single- and double-hump profiles, as in WZ Sge
in quiescence.
These observations have thus shown some degree of resemblance to WZ Sge
in certain aspects.  Since all of the available observations during previous
superoutbursts only covered limited stages of superoutbursts, further
observations of a superoutburst in a more favorable condition, which is
expected to provide a more discriminative information about its suggested
WZ Sge-type nature, have long been waited.  The 1998 November superoutburst
\citep{stu98} was such a long-waited outburst.  The initial detection
by \citet{stu98} was done on 1998 November 10, at a visual magnitude of
11.8, and our observation started within 1 d of the initial detection.
The overall course of the outburst was followed in detail by the present
observation.

\section{Observations}
   The CCD observations were carried out at three observatories: Kyoto,
Tsukuba, and Okayama.  The Kyoto observations were done using an unfiltered
ST-7 camera attached to the Meade 25-cm Schmidt--Cassegrain telescope.
The exposure time was 30 s.  The images were dark-subtracted, flat-fielded,
and analyzed using the Java$^{\rm TM}$-based aperture and PSF photometry
package developed by one of the authors (TK).  The Tsukuba observations
were done using an $R_{\rm c}$ filtered Bitran BT-20 camera attached to
the Meade 25-cm Schmidt--Cassegrain telescope.  The exposure time was
60--180 s.  The images were analyzed using the MIRA A/P aperture photometry
package.  The Okayama observations were made using a $V$-filtered ST-7
camera attached to the Meade 25-cm Schmidt--Cassegrain telescope.  The
exposure time was 15--20 s.  The images were analyzed using a
microcomputer-based aperture photometry package originally developed by
one of the authors (TK) and improved by KM.  All observatories used
GSC 5851.965 (Tycho-2 magnitude $V$=10.59$\pm$0.05, $B-V$=+0.67$\pm$0.09)
as the primary comparison star, whose constancy was confirmed using several
fainter check stars in the same CCD images.  The magnitudes of WX Cet were
determined relative to GSC 5851.965.  Barycentric corrections to observed
times were applied before the following analysis.  A failure of the clock
adjustment was found in the Kyoto November 18 observation.  A retrospective
fine adjustment was made by maximizing the cross-correlation with the
simultaneously taken Tsukuba data.  The zero-point accuracy of the observed
times of the Kyoto November 18 data is thus considered to be $\sim$1 m.

   Since each observer used a different filter, we first added a constant
to each set of observations in order to obtain a common magnitude scale,
which was adjusted to the most abundant Kyoto data.  The constants were
chosen to maximize the cross-correlation after the correction
(table \ref{tab:offset}).  Since outbursting dwarf novae are known to
have colors close to $B-V=0$, the difference in the systems would not
significantly affect the following period analysis.  The log of
observations is given in table \ref{tab:log}.  Nightly averaged magnitudes
listed in the table are corrected by constant offsets in table
\ref{tab:offset}.

\begin{table}
\caption{Magnitude offsets added to each set of observations.}
\label{tab:offset}
\begin{center}
\begin{tabular}{ccc}
\hline\hline
Date (1998) & Okayama data & Tsukuba data \\
\hline
November 11 &    -   & $-$0.011 \\
November 12 &    -   & $-$0.029 \\
November 13 & +0.292 &    -     \\
November 14 & +0.310 &    -     \\
November 15 & +0.337 & +0.051   \\
November 17 &    -   & $-$0.041 \\
November 18 &    -   & $-$0.011 \\
November 19 & +0.199 &    -     \\
November 20 &    -   & $-$0.072 \\
November 24 &    -   & $-$0.121 \\
November 25 &    -   & $-$0.104 \\
\hline\hline
\end{tabular}
\end{center}
\end{table}

\begin{table}
\caption{Log of observations.}\label{tab:log}
\begin{center}
\begin{tabular}{crccrc}
\hline\hline
UT (start--end)  & $N^*$ & Mag$^\dagger$ & Error$^\ddagger$ & Exp$^\|$
                 & Site$^\S$ \\
\hline
1998 November    &     &       &       &     &   \\
11.437 -- 11.591 & 142 & 1.704 & 0.004 & 120 & T \\
11.462 -- 11.592 & 172 & 1.713 & 0.003 &  30 & K \\
12.399 -- 12.656 & 375 & 1.743 & 0.004 &  30 & K \\
12.430 -- 12.690 & 262 & 1.741 & 0.007 &  60 & T \\
13.468 -- 13.703 & 341 & 1.895 & 0.005 &  30 & K \\
13.516 -- 13.557 & 129 & 1.853 & 0.007 &  15 & O \\
14.385 -- 14.697 & 484 & 1.995 & 0.005 &  30 & K \\
14.439 -- 14.561 & 345 & 1.957 & 0.004 &  20 & O \\
15.366 -- 15.677 & 435 & 2.155 & 0.005 &  30 & K \\
15.412 -- 15.646 & 142 & 2.154 & 0.007 & 120 & T \\
15.427 -- 15.579 & 401 & 2.154 & 0.003 &  20 & O \\
17.365 -- 17.700 & 442 & 2.348 & 0.005 &  30 & K \\
17.582 -- 17.660 &  50 & 2.357 & 0.010 & 120 & T \\
18.360 -- 18.713 & 815 & 2.504 & 0.002 &  30 & K \\
18.557 -- 18.668 &  72 & 2.506 & 0.006 & 120 & T \\
19.367 -- 19.707 & 656 & 2.634 & 0.002 &  30 & K \\
19.532 -- 19.657 & 348 & 2.621 & 0.003 &  20 & O \\
20.362 -- 20.648 & 400 & 2.682 & 0.003 &  30 & K \\
20.465 -- 20.479 &  10 & 2.686 & 0.013 & 120 & T \\
22.448 -- 22.688 & 492 & 2.778 & 0.003 &  30 & K \\
23.358 -- 23.687 & 404 & 2.876 & 0.006 &  30 & K \\
24.512 -- 24.687 & 244 & 3.073 & 0.008 &  30 & K \\
24.578 -- 24.655 &  32 & 3.076 & 0.026 & 180 & T \\
25.367 -- 25.678 & 113 & 3.743 & 0.021 &  30 & K \\
25.528 -- 25.646 &  50 & 3.808 & 0.023 & 180 & T \\
26.377 -- 26.680 & 655 & 5.154 & 0.012 &  30 & K \\
29.349 -- 29.657 & 581 & 5.610 & 0.039 &  30 & K \\
December         &     &       &       &     &   \\
 1.350 --  1.576 & 203 & 5.681 & 0.077 &  30 & K \\
 2.388 --  2.613 &  38 & 6.520 & 0.444 &  30 & K \\
 4.397 --  4.441 &   4 & 5.981 & 0.666 &  30 & K \\
 5.365 --  5.652 & 520 & 6.207 & 0.035 &  30 & K \\
 6.372 --  6.627 & 480 & 6.606 & 0.036 &  30 & K \\
 8.498 --  8.654 & 227 & 6.762 & 0.119 &  30 & K \\
 9.351 --  9.641 & 469 & 6.610 & 0.038 &  30 & K \\
10.348 -- 10.607 & 179 & 6.467 & 0.082 &  30 & K \\
\hline
 \multicolumn{6}{l}{$^*$ Number of frames.} \\
 \multicolumn{6}{l}{$^\dagger$ Magnitude relative to GSC 5851.965, corrected} \\
 \multicolumn{6}{l}{\phantom{$^\dagger$} for table \ref{tab:offset}.} \\
 \multicolumn{6}{l}{$^\ddagger$ Standard error of averaged magnitude.} \\
 \multicolumn{6}{l}{$^\|$ Exposure time (s).} \\
 \multicolumn{6}{l}{$^\S$ Sites and filters: T = Tsukuba ($R_{\rm c}$),} \\
 \multicolumn{6}{l}{\phantom{$^\S$} O = Okayama ($V$), K = Kyoto (none).} \\
\end{tabular}
\end{center}
\end{table}

\section{Results}

\subsection{The Outburst Light Curve}

\begin{figure}
  \begin{center}
%    \FigureFile(88mm,60mm){lc.eps}
    \FigureFile(88mm,60mm){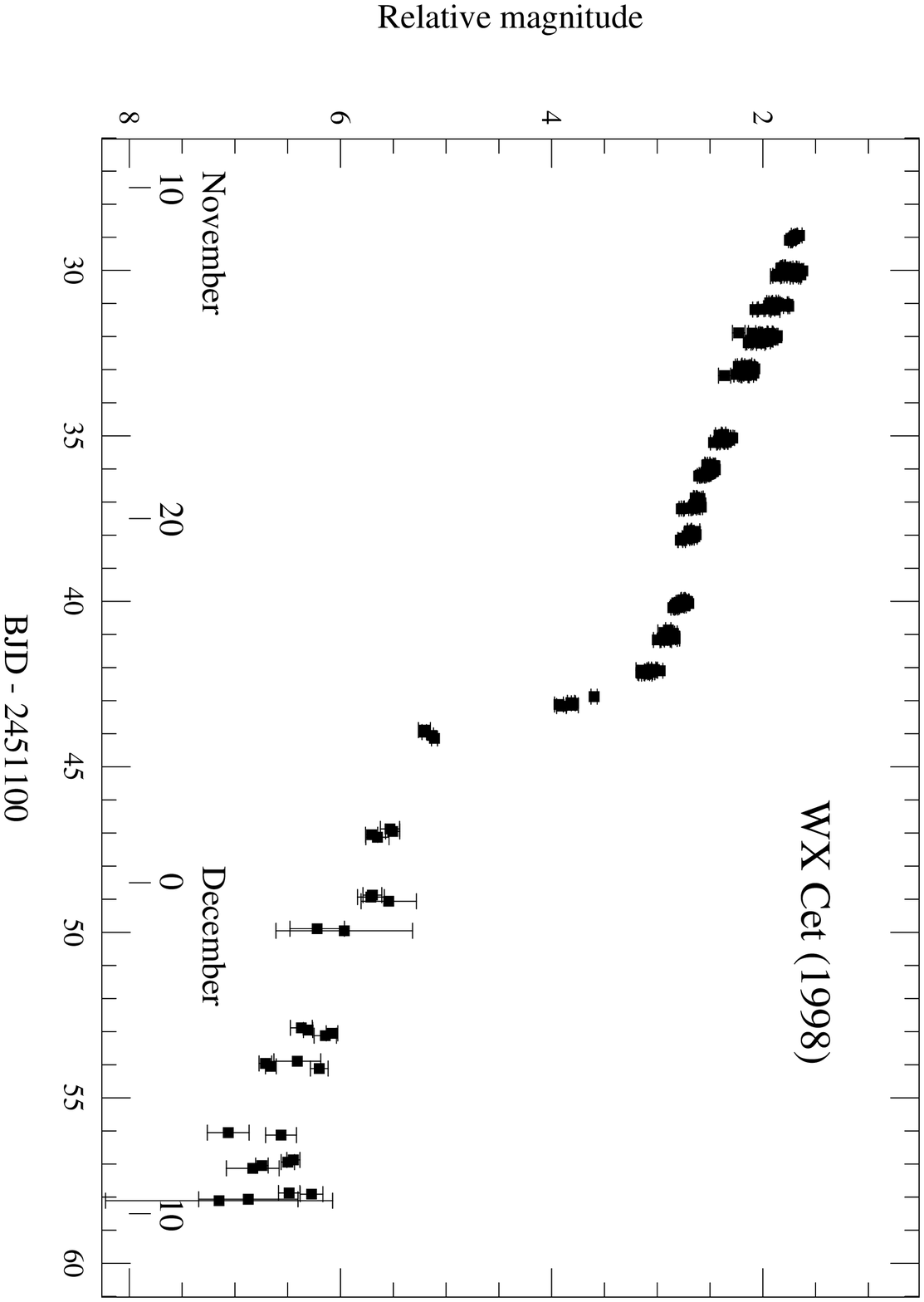}
  \end{center}
  \caption{Light curve of the 1998 superoutburst of WX Cet.  The zero point
  corresponds to magnitude 10.59, on a system close to $R_{\rm c}$.}
  \label{fig:lc}
\end{figure}

   The resultant light curve of the outburst is shown in figure \ref{fig:lc}.
The light curve shows a long-lasting, slowly fading plateau phase, which
is characteristic of an SU UMa-type superoutburst.  The plateau phase
lasted until 1998 November 24 (14 d after the detection of the
outburst\footnote{
  This detection of the outburst can be approximately read as the start
  of the outburst, with an uncertainty of 1 d.
  }).
The object then started fading quickly, and reached the
post-outburst state on November 29, about $\sim$2 mag above the quiescence.
The object showed a gradual fade lasting until the end of the observation
on December 10 (30 d after the detection of the outburst).
Such a long, fading tail is considered as a relatively common, but not
always exclusive, feature of WZ Sge-type stars \citep{kat97}.

\subsection{Superhumps}\label{sec:sh}

\begin{figure}
  \begin{center}
%    \FigureFile(80mm,60mm){early.eps}
    \FigureFile(80mm,60mm){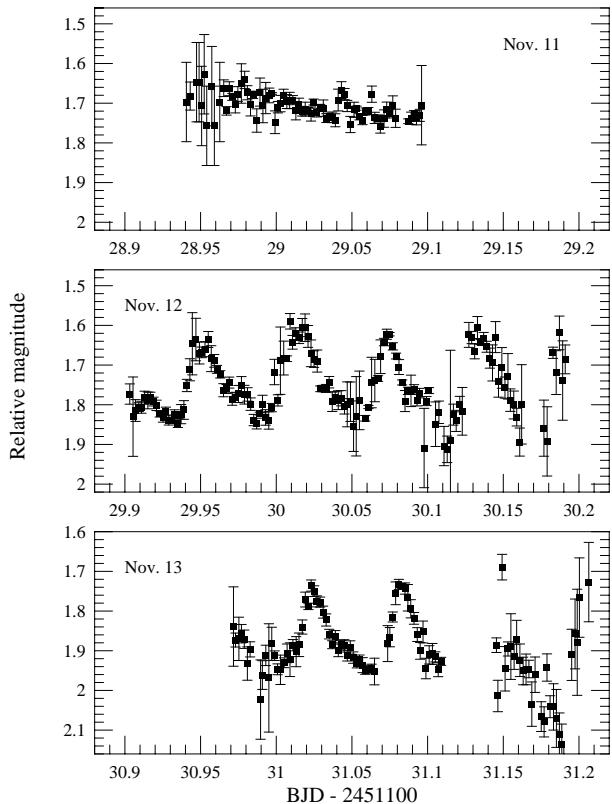}
  \end{center}
  \caption{First three nights of the 1998 superoutburst of WX Cet.
  Prominent superhumps appeared on the second night of the observation
  (1998 November 12).}
  \label{fig:early}
\end{figure}

   Figure \ref{fig:early} shows an enlargement of the early part of the
light curve.  On the first night (November 11), the object showed little
variation.  On the next night (November 12), prominent superhumps
developed.

\begin{figure}
  \begin{center}
%    \FigureFile(70mm,90mm){pdm.eps}
    \FigureFile(70mm,90mm){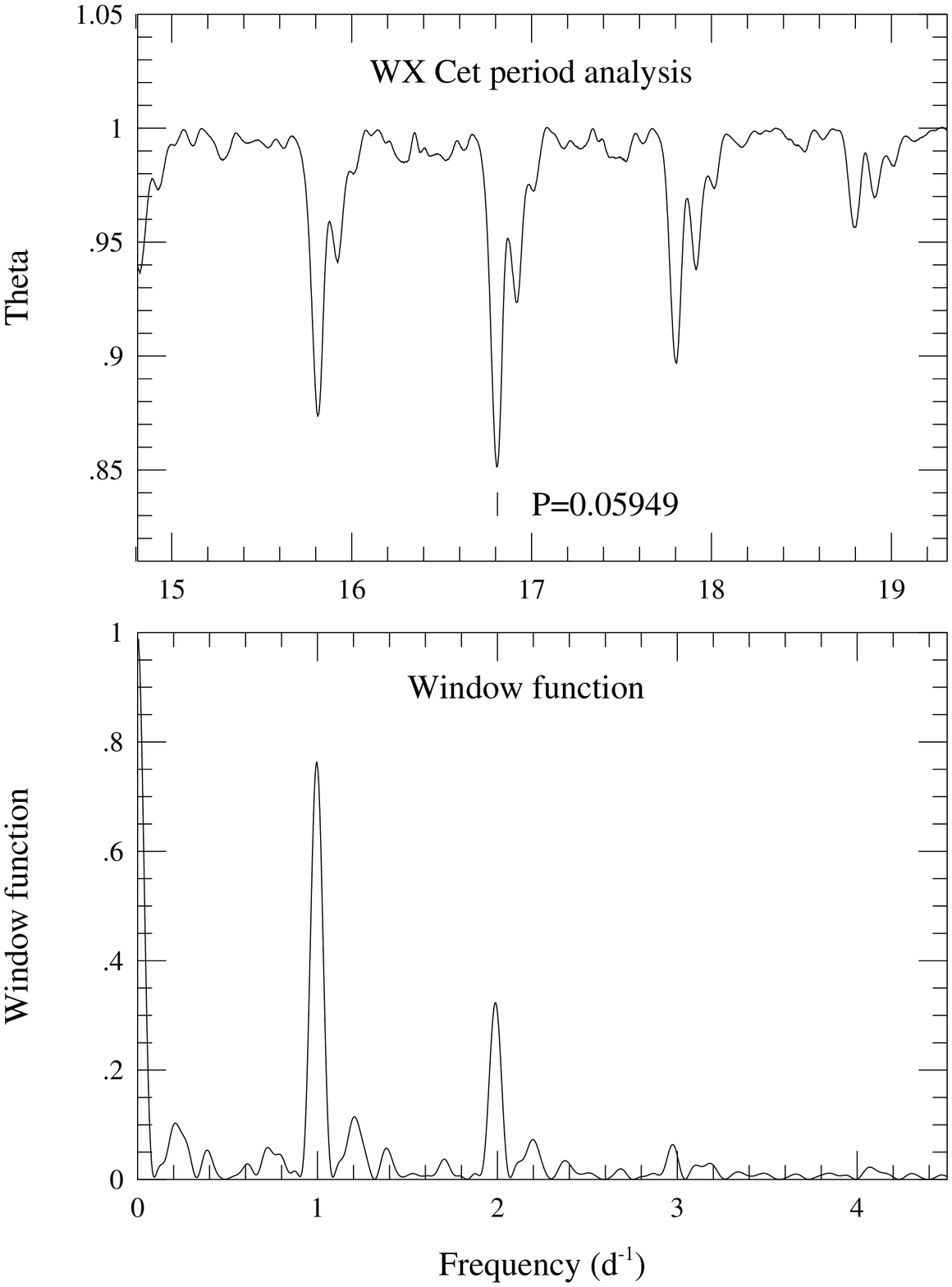}
  \end{center}
  \caption{Period analysis of superhumps in WX Cet.  The analysis was
  done for the data between 1998 November 12 and 25, when superhumps
  were prominently observed.  (Upper panel) The Phase Dispersion
  Minimization (PDM) method \citep{ste78} was used after removing the
  slow trend of decline and prewhitening for slow variations with
  frequencies smaller than 2 d$^{-1}$.  The signal at the frequency
  16.81 d$^{-1}$ corresponds to a mean superhump period of 0.05949 d.
  (Lower panel) The window function.  A slight asymmetry of the signal
  in the PDM analysis is mostly attributable to the asymmetry of the
  window function.
  }
  \label{fig:pdm}
\end{figure}

\begin{figure}
  \begin{center}
%    \FigureFile(80mm,60mm){phaseave.eps}
    \FigureFile(80mm,60mm){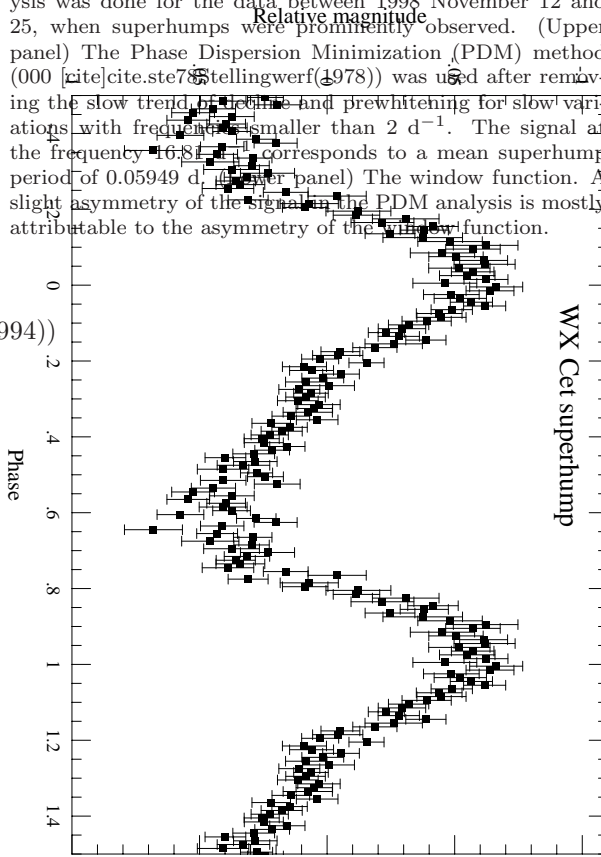}
  \end{center}
  \caption{Phase-averaged light curve of WX Cet superhumps.}
  \label{fig:phaseave}
\end{figure}

   We applied the Phase Dispersion Minimization (PDM) method \citep{ste78}
to all of the data between 1998 November 12 and 25, after removing the
slow trends of decline, and after prewhitening for slow variations with
frequencies smaller than 2 d$^{-1}$.  The resultant theta diagram is shown
in figure \ref{fig:pdm}.  The signal at the frequency 16.81 d$^{-1}$
corresponds to the mean superhump period ($P_{\rm SH}$) of 0.05949(1) d.
A slight asymmetry of the signal in the PDM analysis is mostly attributable
to the asymmetry of the window function.
The resultant superhump period gives a fractional superhump excess
($\epsilon=P_{\rm SH}/P_{\rm orb}-1$) of 2.1\% against \citet{men94} and
\citet{rog01}.  Figure \ref{fig:phaseave} shows the averaged profile
of superhumps.  The phase zero is taken as BJD 2451129.952 (1998 November
12.452 UT).

\begin{figure}
  \begin{center}
%    \FigureFile(88mm,110mm){nightave.eps}
    \FigureFile(88mm,110mm){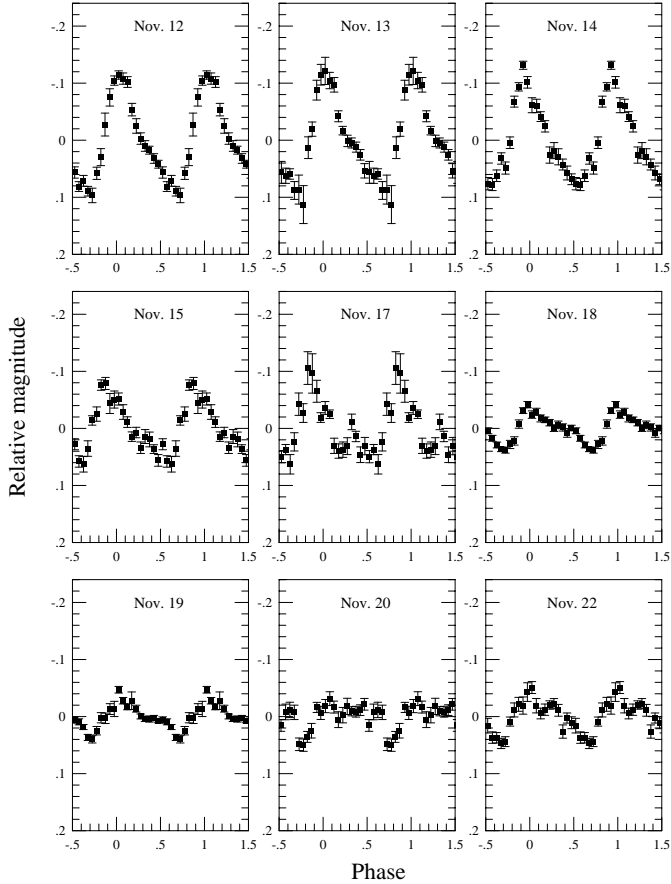}
  \end{center}
  \caption{Evolution of WX Cet superhumps.}
  \label{fig:nightave}
\end{figure}

   Figure \ref{fig:nightave} shows nightly averaged profiles of superhumps.
The superhumps quickly grew in amplitude, and slowly decayed.  The shifts
of the superhump maxima toward negative phases represent the $O-C$
variation described in subsection \ref{sec:oc}.  The superhumps became less
prominent after $\sim$ 8d of the appearance of superhumps, but the
amplitude grew again near the terminal stage of the superoutburst plateau.
Such a regrowth of superhumps at the later stage was also observed in a
large-amplitude SU UMa-type dwarf nova, V1028 Cyg, which \citet{bab00}
considered as an intermediate object between usual SU UMa-type dwarf novae
and WZ Sge-type stars.

\subsection{Early Superhumps?}\label{sec:earlysh}

\begin{figure}
  \begin{center}
%    \FigureFile(70mm,55mm){ph11.eps}
    \FigureFile(70mm,55mm){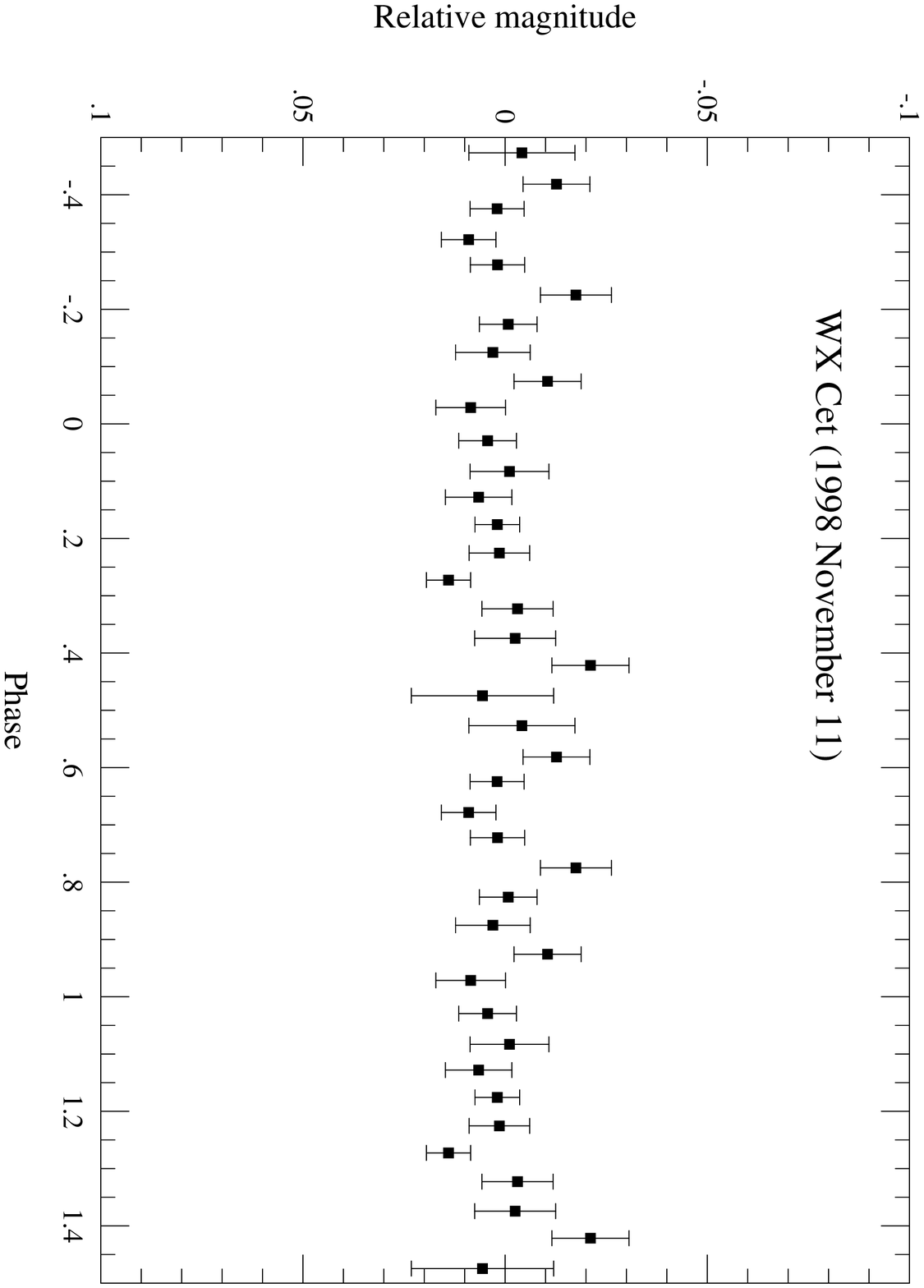}
  \end{center}
  \caption{Phase-averaged light curve of November 11 at the reported
  $P_{\rm orb}$.  The phase was taken arbitrarily.  The upper limit of
  ``early superhump"-type modulations is 0.03 mag.}
  \label{fig:ph11}
\end{figure}

   All of the superoutbursts of well-observed WZ Sge-type dwarf novae
show semi-periodic modulations at the earliest stage of a superoutburst
(\cite{kat96}; \cite{mat98}; \cite{kat98}), which have the same, or
extremely close, periods to $P_{\rm orb}$.  These modulations are
called ``early superhumps" (e.g. \cite{kat98}
  \footnote{
  This feature is also referred to as ``orbital" superhumps or
  {\it outburst orbital hump} \citep{patetal98}.
  }).
Although the origin of these modulations is not still perfectly understood,
the exertion of a tidal instability on the accretion disk reaching the
3:1 resonance (the resonance responsible for the tidal instability)
during the long quiescent states of WZ Sge-type stars is a promising
explanation (\cite{mey98}; \cite{min98}).  The presence of early
superhumps can thus be considered to be one of the photometric criteria for
WZ Sge-type stars.
The data on November 11 (the first day of the observation) were examined.
After removing the linear trend, the data were analyzed using the PDM.
No significant signals were detected close to $P_{\rm orb}$ or
$P_{\rm SH}$.  Figure \ref{fig:ph11} shows the phase-averaged light curve
of the November 11 data at the reported $P_{\rm orb}$.  No clear periodic
signal was detected at this period.  The upper limit of
``early superhump"-type modulations was 0.03 mag.

\subsection{$O-C$ Changes}\label{sec:oc}

   We extracted the maxima times of superhumps from the light curve by eye.
The averaged times of a few to several points close to the maximum were
used as representatives of the maxima times.  Thanks to high signal-to-noise,
densely sampled data, the errors of the maxima times are usually less than
$\sim$0.002 d.  The resultant superhump maxima are given in table
\ref{tab:shmax}.  The values are given to 0.0001 d in order to avoid
the loss of significant digits in a later analysis.  The cycle count
($E$) is defined as the cycle number since BJD 2451129.953 (1998 November
12.453 UT).  A linear regression to the observed superhump times,
disregarding late superhumps (discussed in the next \ref{sec:latesh}),
gives the following ephemeris:

\begin{equation}
{\rm BJD (maximum)} = 2451129.9502 + 0.0594945 E. \label{equ:reg1}
\end{equation}

\begin{table}
\caption{Times of superhump maxima.}\label{tab:shmax}
\begin{center}
\begin{tabular}{ccc}
\hline\hline
$E^*$  & BJD$-$2400000 & $O-C^\dagger$ \\
\hline
  0    & 51129.9531 &  0.0029 \\
  1    & 51130.0139 &  0.0042 \\
  2    & 51130.0744 &  0.0052 \\
  3    & 51130.1352 &  0.0066 \\
 18    & 51131.0242 &  0.0031 \\
 19    & 51131.0822 &  0.0016 \\
 34    & 51131.9707 & -0.0023 \\
 35    & 51132.0312 & -0.0013 \\
 36    & 51132.0916 & -0.0004 \\
 37    & 51132.1482 & -0.0033 \\
 50    & 51132.9176 & -0.0073 \\
 51    & 51132.9781 & -0.0063 \\
 52    & 51133.0346 & -0.0093 \\
 86    & 51135.0630 & -0.0037 \\
 87    & 51135.1195 & -0.0067 \\
 88    & 51135.1808 & -0.0049 \\
100    & 51135.9018 &  0.0022 \\
101    & 51135.9588 & -0.0003 \\
102    & 51136.0189 &  0.0003 \\
103    & 51136.0801 &  0.0020 \\
104    & 51136.1383 &  0.0007 \\
119    & 51137.0316 &  0.0016 \\
120    & 51137.0930 &  0.0035 \\
121    & 51137.1502 &  0.0012 \\
134    & 51137.9292 &  0.0068 \\
135    & 51137.9879 &  0.0060 \\
169    & 51140.0038 & -0.0009 \\
170    & 51140.0654 &  0.0012 \\
171    & 51140.1223 & -0.0014 \\
172    & 51140.1817 & -0.0015 \\
187    & 51141.0718 & -0.0038 \\
188    & 51141.1343 & -0.0008 \\
204    & 51142.0942 &  0.0072 \\
205    & 51142.1442 & -0.0023 \\
235$^\ddagger$ & 51143.9447 & 0.0133 \\
236$^\ddagger$ & 51144.0074 & 0.0165 \\
237$^\ddagger$ & 51144.0880 & 0.0376 \\
238$^\ddagger$ & 51144.1466 & 0.0367 \\
\hline
 \multicolumn{3}{l}{$^*$ Cycle count since BJD 2451129.953.} \\
 \multicolumn{3}{l}{$^\dagger$ $O-C$ calculated against equation
                    \ref{equ:reg1}.} \\
 \multicolumn{3}{l}{$^\ddagger$ Late superhumps.} \\
\end{tabular}
\end{center}
\end{table}

\begin{figure}
  \begin{center}
%    \FigureFile(80mm,60mm){oc.eps}
    \FigureFile(80mm,60mm){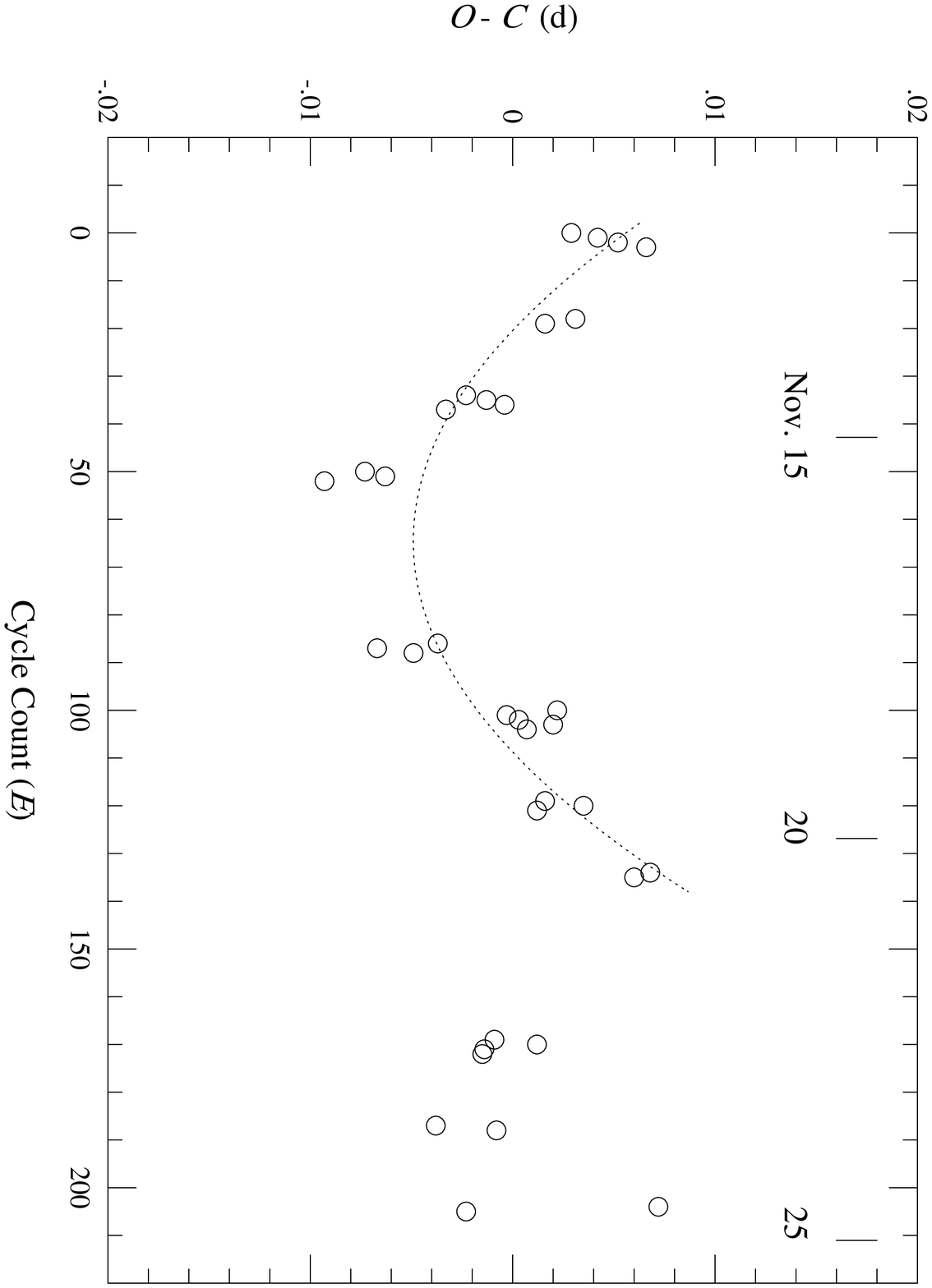}
  \end{center}
  \caption{$O-C$ diagram of superhump maxima.
  The parabolic fit corresponds to equation \ref{equ:reg2}.}
  \label{fig:oc}
\end{figure}

   Figure \ref{fig:oc} shows the ($O-C$)'s against the mean superhump period
(0.05949 d).  The diagram clearly shows the increase in the superhump
period between $E=0$ (1998 November 12) and $E=140$ (November 20, 10 d
after the outburst detection).  This interval corresponds to the
superoutburst plateau.  The times of the superhump maxima in this interval
can be well represented by the following quadratic equation:

\begin{eqnarray}
{\rm BJD} & {\rm (maximum)} = 2451129.9558(10) + 0.059169(39) E \nonumber \\
    & + 2.52(29) \times 10^{-6} E^2. \label{equ:reg2}
\end{eqnarray}

   The quadratic term corresponds to $\dot{P}$ = +5.0$\pm$0.6 $\times$
10$^{-6}$ d cycle$^{-1}$, or $\dot{P}/P$ = +8.5$\pm$1.0 $\times$ 10$^{-5}$,
which is one of the largest period derivatives ever observed in all
SU UMa-type dwarf novae \citep{kat98}.

\subsection{Late Superhumps}\label{sec:latesh}

\begin{figure}
  \begin{center}
%    \FigureFile(88mm,110mm){latesh.eps}
    \FigureFile(88mm,110mm){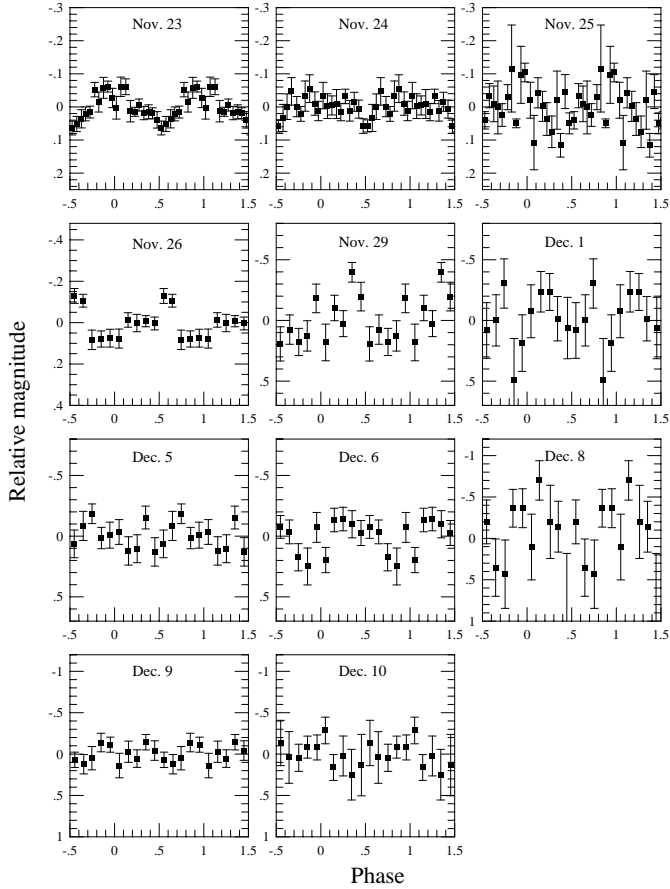}
  \end{center}
  \caption{Phase-averaged light curve of WX Cet superhumps, after the final
  stage of the superoutburst.}
  \label{fig:latesh}
\end{figure}

   During the final stage of a superoutburst and the subsequent
post-superoutburst stages, some SU UMa-type dwarf nova show modulations
having approximately the same period as $P_{\rm orb}$, but having a maximum
phase of $\sim$0.5 offset from those of usual superhumps.  This phenomenon
is called ``late superhumps" (\cite{hae79}; \cite{vog83};
\cite{vanderwoe88}).  Figure \ref{fig:latesh} shows the nightly averaged
profiles of variations.  The time of phase zero and the period
($P_{\rm SH}$) used in folding are the same as in figure
\ref{fig:nightave}.  Note that the vertical scales are different between
the panels.  The relatively large error bars are a result of the faintness
of the object.  Superhumps at normal phases (around phase = 0)
persisted on November 23, but became weaker on the next night.
The signal became slightly stronger on November 25 (the rapid decline
stage).  A phase reversal was clearly observed on November 26,
corresponding to the appearance of late superhumps.  The signal
likely persisted until November 29, but became weaker and more
irregular on subsequent nights.

\section{Discussion}

\subsection{Superhump Period Excess}

   The present observation provides a superhump period of
0.05949(1) d (average), best determined in the long research
history of WX Cet.  This period corresponds to a fractional superhump
excess of ($\epsilon$) 2.1\%.  This value is moderately small among SU UMa
stars (cf. \cite{pat98}), but is substantially larger than those of
some of well-established WZ Sge-type dwarf novae: WZ Sge, $\epsilon$ = 0.8\%
\citep{pat81}; AL Com, $\epsilon$ = 1.0\% (cf. \cite{nog97} for a
comprehensive summary). Some WZ Sge-type dwarf novae (HV Vir,
$\epsilon$ = 2.0\% \citep{kat01}; EG Cnc [controversy exists; \citet{kat97}
gives $\epsilon$ = 2.7\% while \citet{patetal98} suggests $\epsilon$=0.7\%],
however, seem to have similar $\epsilon$ to that of WX Cet.
Both analytical analysis (e.g. \cite{osa85}) and numerical simulations
(e.g. \cite{mur98}) suggest that $\epsilon$ is a good measure of the
binary mass ratio $q$.  With this regard, WX Cet is expected to have an
intermediate binary parameters between (extreme) WZ Sge-type dwarf novae
and SU UMa-type dwarf novae.

\subsection{Early Superhumps}

   As described in subsection \ref{sec:earlysh}, the existence of early
superhumps is one of the diagnostic features of WZ Sge-type dwarf novae.
In the present observation of WX Cet, no evidence of early superhumps
was observed.  Even if the true maximum of the present superoutburst
was missed by a day (only 1-d observational gaps existed before the
outburst detection), the disappearance of early superhumps within two
day from the start of the outburst makes a clear contrast to the week-long
persistence of early superhumps in a WZ Sge-type star, AL Com (Ishioka
et al., in preparation).  Both the lack of a clear signal of early
superhumps and the quick evolution of usual superhumps
(subsection \ref{sec:sh}) are against the interpretation of WX Cet as
a WZ Sge-type dwarf nova.  A similar evolution of superhumps was
also observed in V1028 Cyg \citep{bab00}, a star which showed intermediate
properties between the usual SU UMa-type dwarf novae and WZ Sge-type stars.

    As introduced in subsection \ref{sec:earlysh}, recent theoretical studies
suggest that the work of tidal instability during the long quiescence of
WZ Sge-type dwarf novae is essential for the appearance of early superhumps.
It is not surprising that early superhumps are the most discriminative
feature of WZ Sge-type dwarf novae, since the long-lasting, large-amplitude
superoutbursts of WZ Sge-type dwarf novae are regarded as a necessary
consequence of a combination of low mass-transfer rates and the effective
removal of the innermost accretion disk, which leads to an expansion of
the accretion disk, eventually reaching the tidal instability (\cite{mey98};
\cite{min98}).  From the lack of a clear signature of early superhumps,
we consider WX Cet to be a fairly normal SU UMa-type dwarf nova with
a large outburst amplitude, rather than a WZ Sge-type dwarf nova.

\subsection{Period Changes}

   As shown in subsection \ref{sec:oc}, WX Cet showed a clear increase
of the superhump period during the plateau stage of the superoutburst.
Only a limited number of SU UMa-type dwarf novae are known to have
such positive period derivatives (cf. \cite{kat98}; \cite{bab00}
and references therein).  Since most of these systems are short
$P_{\rm orb}$ systems, there has been a suggestion that low $q$ and/or
low $\dot{M}$ are responsible for the phenomenon \citep{kat98}.
Recent discoveries of zero to marginally positive $\dot{P}$ systems
(V725 Aql: \cite{uem01}; EF Peg: Matsumoto et al, in preparation)
in long $P_{\rm orb}$ systems more support that low $\dot{M}$
is more responsible for the phenomenon.  [Note, however, the peculiar,
relatively high $\dot{M}$ system V485 Cen is also known to show a
positive $\dot{P}$ \citep{ole97}.  This may be an indication of
the limit of our understanding of this phenomenon].
Although a positive $\dot{P}$ is frequently met in WZ Sge-type systems,
this may not be considered as a diagnostic feature.

\subsection{Late Superhumps}

   In recent examples of well-observed WZ Sge-type stars (EG Cnc and AL Com),
late superhumps (subsection \ref{sec:latesh}) were observed to persist for
more than tens of days after the termination of the main superoutburst
(\cite{kat97}; \cite{patetal98}; \cite{nog97}).  Although the detection
in the present observation was limited because of the faintness of the
object, a clear decay of late superhumps within 4 d of the termination of
the superoutburst (figure \ref{fig:latesh}) suggests that large-amplitude
late superhumps rather quickly decayed in this system.  The time scale of
the decay is roughly comparable to those in usual SU UMa-type dwarf novae
\citep{hae79}.  The origin of late superhumps is proposed to be
modulation of the properties of a precessing accretion disk at the stream
impact point \citep{hes92}.  While it is not still clear why this effect
persists longer in WZ Sge-type dwarf novae, a low $\dot{M}$ may be responsible
for delaying circularizing of the disk, resulting in a persistence
of the precessing accretion disk.  This may not be a direct discriminating
feature of WZ Sge-type dwarf novae, but apparently needs to be examined
in larger samples.

\subsection{Quiescent Humps}

   \citet{rog01} noted that the orbital light curve showed alternations
between single- and double-hump profiles.  \citet{rog01} discussed the
resemblance of this phenomenon with the similar alternations of humps
in quiescent WZ Sge.  Since the present work more strongly supports that
WX Cet is a rather normal SU UMa-type dwarf nova with a large outburst
amplitude, rather than an extreme WZ Sge-type star, such alternations
of hump profiles in quiescence may not be a discriminative feature of
WZ Sge-type dwarf novae.  The presence of double humps may be better
explained by the ``reversed hot spot" hypothesis by \citet{men94}
[see also \citet{men99} for an example of changing hump profiles in
a low $\dot{M}$ dwarf nova].

\subsection{Rebrightening and the Late Decay Stage}

   There was no indication of a post-superoutburst rebrightening, which
is often associated in short $P_{\rm orb}$ SU UMa-type dwarf novae,
especially in WZ Sge-type dwarf novae (cf. \cite{kat98}), both in
our observations and in visual observations reported to the VSNET
Collaboration\footnote{
  $\langle$http://www.kusastro.kyoto-u.ac.jp/vsnet/$\rangle$.
}.

   The origin of post-superoutburst rebrightening in large-amplitude
SU UMa-type dwarf novae was proposed to be a reflection of cooling wave
in the accretion disk \citep{how95}.  A more recent explanation includes
the work by \citet{osa97}, who assumed that the quiescent viscosity of the
accretion disk is somehow maintained higher following a superoutburst
than in quiescence.  \citet{osa01} further succeeded in reproducing
a variety of post-superoutburst rebrightenings -- from no rebrightening
to multiple rebrightenings -- by considering the competition between
the thermal disk instability and the recently discovered mechanism of
a decay of MHD turbulence under the condition of the low magnetic Reynolds
numbers in the cold accretion disk \citep{gam98}.  WX Cet showed a slow
fading after the superoutburst (figure \ref{fig:lc}).  This phenomenon may
be an exemplification of a gradual decay of the disk viscosity, which
determines the luminosity of the disk.  Following \citet{osa01}, we suspect
that the ignition of thermal instability accidentally failed to occur
under competition with the decay of viscosity in the present
post-superoutburst state of WX Cet.  The post-superoutburst decline of
WX Cet was almost perfectly linear (exponential), with a rate of
0.10 mag d$^{-1}$.  Interestingly, this rate of decline is almost
perfectly identical with the mean rate of decline (0.10 mag d$^{-1}$)
during the plateau stage.
Although this coincidence may be merely accidental, this may suggest the
existence of a time scale for the decay of the disk viscosity related to
the decay of the superoutburst plateau.

\section{Conclusion}

   We observed the 1998 November superoutburst of WX Cet, a dwarf nova
which had been proposed to be a related system to the peculiar dwarf nova
WZ Sge.  The observation established that WX Cet is an SU UMa-type
dwarf nova with a mean superhump period of 0.05949(1) d.
This period is 2.1\% longer than the reported orbital period.
This fractional superhump excess lies between those of extreme
WZ Sge-type systems and those of other SU UMa-type systems.
The lack of early superhumps at the earliest stage of superoutburst,
together with the rapid development of usual superhumps, seems to
disqualify WX Cet as being a WZ Sge-type dwarf nova.  A period increase
of superhumps with $\dot{P}/P$ = +8.5$\pm$1.0 $\times$ 10$^{-5}$ was
observed during the superoutburst plateau.  This is one of the largest
$\dot{P}/P$ ever observed in SU UMa-type dwarf novae.  Although there was
no evidence of a post-superoutburst rebrightening, a linear decline,
with a rate of 0.10 mag d$^{-1}$, was observed in the post-superoutburst
stage.  This may be an exemplification of the proposed decay of quiescent
viscosity following the superoutburst.

\vskip 3mm

We are grateful to many amateur observers for supplying their vital visual
and CCD estimates via VSNET, and especially to Rod Stubbings for his
detection and early notification of the long-awaited superoutburst.
Part of this work is supported by a Research Fellowship of the
Japan Society for the Promotion of Science for Young Scientists (KM).

\end{document}